\documentclass[preprint,12pt]{elsarticle}

\usepackage{amssymb}
\usepackage{amsmath}
\biboptions{numbers,sort&compress}

\def\lsi{\raise0.3ex\hbox{$<$\kern-0.75em\raise-1.1ex\hbox{$\sim$}}}
\def\gsi{\raise0.3ex\hbox{$>$\kern-0.75em\raise-1.1ex\hbox{$\sim$}}}

\newcommand{\fig}{Fig.~}

\newcommand{\bea}{\begin{eqnarray}}
\newcommand{\eea}{\end{eqnarray}}
\newcommand{\beq}{\begin{equation}}
\newcommand{\eeq}{\end{equation}}
\newcommand{\bd}{\begin{displaymath}}
\newcommand{\ed}{\end{displaymath}}
\newcommand{\bpm}{\begin{pmatrix}}
\newcommand{\epm}{\end{pmatrix}}
\newcommand{\nn}{\nonumber}

\def\bx{{\mathbf{x}}}

\def\bp{{\mathbf{p}}}
\def\bu{{\mathbf{u}}}

\journal{Journal of Subatomic Particles and Cosmolgy}

\begin{document}

\begin{frontmatter}



\title{Spectral properties of pseudo-scalar mesons through the QCD chiral crossover}


\author[2]{P.~Lowdon}
\author[2,3]{O.~Philipsen}

\affiliation[2]{organization={ITP, Goethe-Universität Frankfurt am Main},
            addressline={Max-von-Laue-Str. 1}, 
            city={60438 Frankfurt am Main},
            country={Germany}}
\affiliation[3]{organization={John von Neumann Institute for Computing (NIC) at GSI},
            addressline={Planckstr. 1}, 
            city={64291 Darmstadt},
            country={Germany}}

\begin{abstract}
We summarise recent progress towards the non-perturbative determination of thermal spectral functions for pseudo-scalar mesons
in QCD by exploiting constraints imposed by micro-causality at finite temperature. For temperatures not much above the
vacuum particle mass, continuous contributions from scattering, Landau damping and collective excitations are found to be
negligible. This allows for a quantitative description of spatial and temporal lattice correlators
in terms of thermoparticles, i.e.~vacuum excitations modified by medium effects, with resonance-like structures persisting
for a range above the chiral crossover.
\end{abstract}





\end{frontmatter}

\section{Introduction: Spectral functions in vacuum QFT}
\label{sec1}

Spectral functions provide an understanding of the effective degrees of freedom in a
quantum field theory (QFT), both in vacuum and in medium, and furnish a bridge from theory to experiment. 
This is particularly relevant for QCD, whose dynamics
is expected to change from hadronic to partonic depending on the scales set by temperature or density, where
any change of dynamics is encoded in the same gauge-invariant correlation functions.

\begin{figure}[t]
\centering
\includegraphics[width=0.6\textwidth]{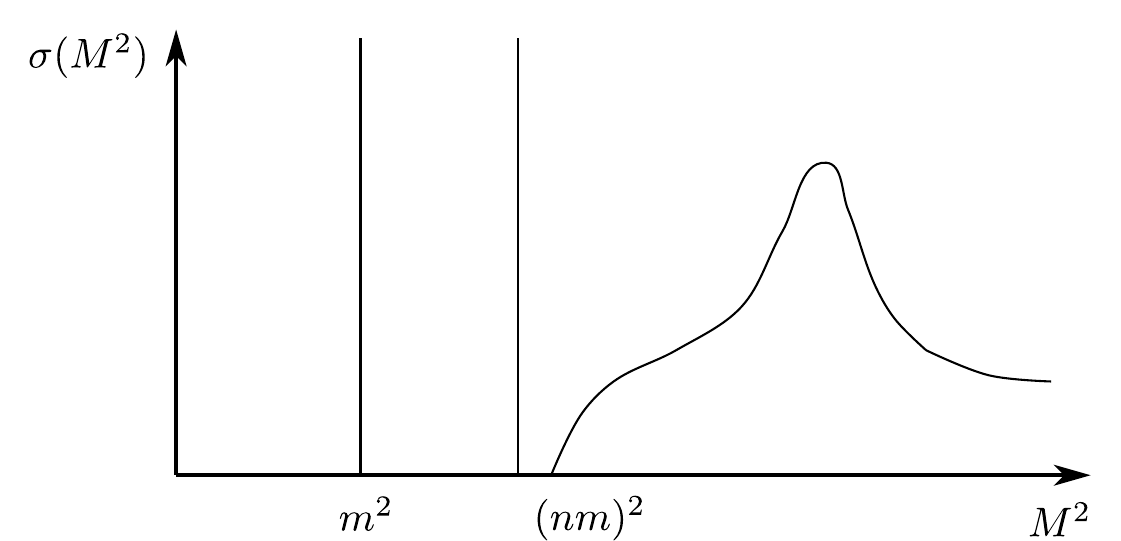}
\caption{Schematic spectral function, with delta functions representing discrete particle (elementary or bound) states, 
and a continuum of scattering states including resonances. }
\label{fig1}
\end{figure}
Let us recall the definition of the spectral function for a real scalar field as the Fourier transform of the expectation value of the field commutator,
\beq
   \langle \Omega | \; \Big[\phi(x), \phi(y) \Big]\;|\Omega\rangle 
\equiv \int\!\frac{d^4p}{(2\pi)^4} \;\rho(p) e^{-ip(x-y)}. \nn
\label{eq:caus3}
\eeq
In a relativistic QFT micro-causality ensures $\langle \Omega| [\phi(x),\phi(y)]|\Omega\rangle=0$ for space-like separations $(x-y)^2<0$. 
By its definition the spectral function thus expresses micro-causality in momentum space.    
The formal solution is
\beq
\rho(p)=\sigma(p^2)\varepsilon(p^0),\qquad
\sigma(p^2)=2\pi \sum_\lambda  | \langle \Omega|\phi(0)|\lambda_0\rangle |^2 \delta(p^2-m_\lambda^2).
\label{eq:sigma}
\eeq
Here, $\lambda$ labels the eigenstates of the Hamiltonian, including multi-particle states, and $\lambda_0$ denotes
zero momentum for their respective center of mass. Now the entire spectrum of the theory and their
matrix elements in the last equation are required, which are not accessible analytically.
The full two-point function can be expressed in terms of the spectral function and the Feynman propagator with mass $M$, known
as K\"allen-Lehman representation, 
\beq
C(x-y)=\langle \Omega|T\Big(\phi(x)\phi(y)\Big)|\Omega\rangle \nn\\
=
\int_0^\infty d M^2\; \sigma(M^2) \Delta_F(x-y;M^2). 
\label{eq:specdef}
\eeq
Conversely this means that the entire spectrum of the theory, consisting of single particle, bound and scattering states
 as shown schematically in \fig\ref{fig1},
contributes and is related to the two-point function.

\section{Spectral functions and thermoparticles at finite temperature}

For finite temperature equilibrium physics we consider our QFT of interest at Euclidean times $\tau=-it$. For many 
practical applications we are interested in the partially Fourier transformed two-point function
\begin{align}
C(\tau,\bp) =
 \int_{0}^{\infty} \frac{d\omega}{2\pi}  \,\frac{\cosh(\omega(|\tau|-\beta/2))}{\sinh(\beta\omega/2)} \rho(\omega,\bp).
\end{align}
For QCD such correlators can be calculated straightforwardly on the lattice, but extracting the spectral function from 
a discrete set of data points with statistical errors amounts to an ill-defined inversion problem. 
Here we use an approach employing that, 
at finite temperature, an additional constraint  besides micro-causality is the KMS condition, i.e.~periodicity
of correlators in Euclidean time. Together these imply the following representation of the spectral 
function at finite temperature~\cite{Bros:1992ey,Bros:1996mw,Bros:1998ua}, $\beta=1/T$,
\begin{align}
\rho(\omega,\bp) = \int_{0}^{\infty} \! ds \int \! \frac{d^{3}\bu}{(2\pi)^{2}} \ \epsilon(\omega) \, \delta\!\left(\omega^{2} - (\bp-\bu)^{2} - s \right)\widetilde{D}_{\beta}(\bu,s),
\label{eq:commutator_rep}
\end{align}
generalising the K\"allen-Lehman representation to QFT at finite temperature.

In the following we employ lattice simulation results of on-axis spatial and temporal correlators, 
\beq
C^s(z)=\sum_{x,y,\tau} C(\tau,\bx)\label{eq:c_z},\quad
C^\tau(\tau)=\sum_{x,y,z}C(\tau,\bx).
\eeq
The most direct route to the thermal spectral density in an isotropic medium is via the spatial correlator and the relation~\cite{Lowdon:2022xcl} 
\begin{align}
C^s(z) = \int^{\infty}_{-\infty} \! dx \, dy  \int_{0}^\beta d\tau \, C(\tau, \bx)  
  = \frac{1}{2}\int_{0}^{\infty} \! ds \int^{\infty}_{|z|} \! dR \ e^{-R\sqrt{s}} D_{\beta}(R,s).
\label{C_int}
\end{align} 
In this form we still have an inversion problem. However, we can now make use of a natural decomposition of 
the thermal spectral density for stable spinless particles in terms of the ansatz~\cite{Bros:2001zs}
\begin{align}
D_{\beta}(\bx,s)= \sum_{i} D_{m_{i},\beta}(\bx)\, \delta(s-m_{i}^{2}) + D_{c, \beta}(\bx,s).
\label{D_decomp}
\end{align}
The first term constitutes the so-called thermoparticle component. It consists of a discrete sum over  
the stable vacuum particles, which contribute with their on-shell condition 
modified by a damping factor representing the medium effects. In the limit
$T\rightarrow 0$ the damping factor approaches unity and the first term reduces to the stable particle contribution to
the vacuum spectral function. The second term contains all continuous contributions such as scattering states, collective
excitations and Landau damping, with the latter two vanishing for $T\rightarrow 0$. For massive particles there is a gap between the one-particle and scattering states.
Thus at  sufficiently low temperatures, the thermoparticle components may be expected to dominate the correlator 
according to (\ref{C_int}), so that dropping the continuous part should be a reasonable (and testable) approximation. In this
case the spatial correlator simplifies to
\begin{align}
C(z) \approx \frac{1}{2}\sum_{i=1}^{n} \,  \int^{\infty}_{|z|} \! dR \  e^{-m_{i}R} D_{m_{i},\beta}(R),
\label{C_decomp_dom}
\end{align}
and the damping factor can be computed from the correlator via
\begin{align}
D_{m_{1},\beta}(|\bx|=z) \sim -2e^{m_{1} |z|} \, \frac{d C(z)}{dz}, \quad\quad z \rightarrow \infty.
\label{D_C_rel}
\end{align}
\begin{figure}[t]
\centering
\includegraphics[width=0.45\textwidth]{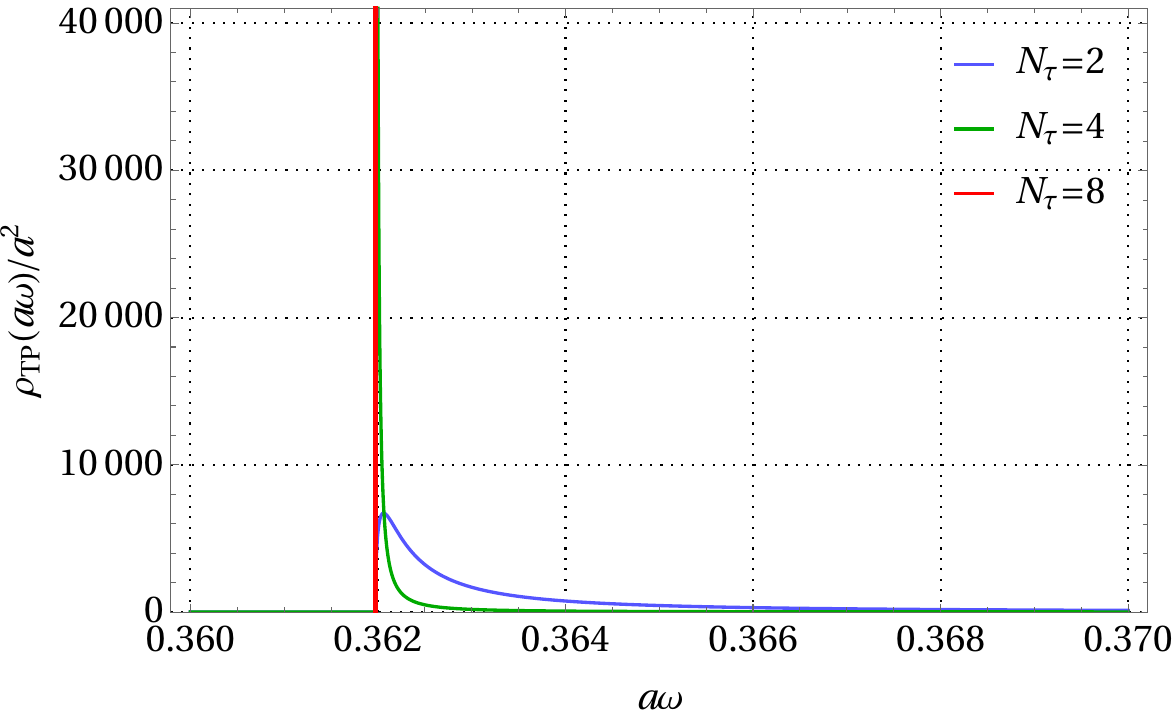}
\includegraphics[width=0.42\textwidth]{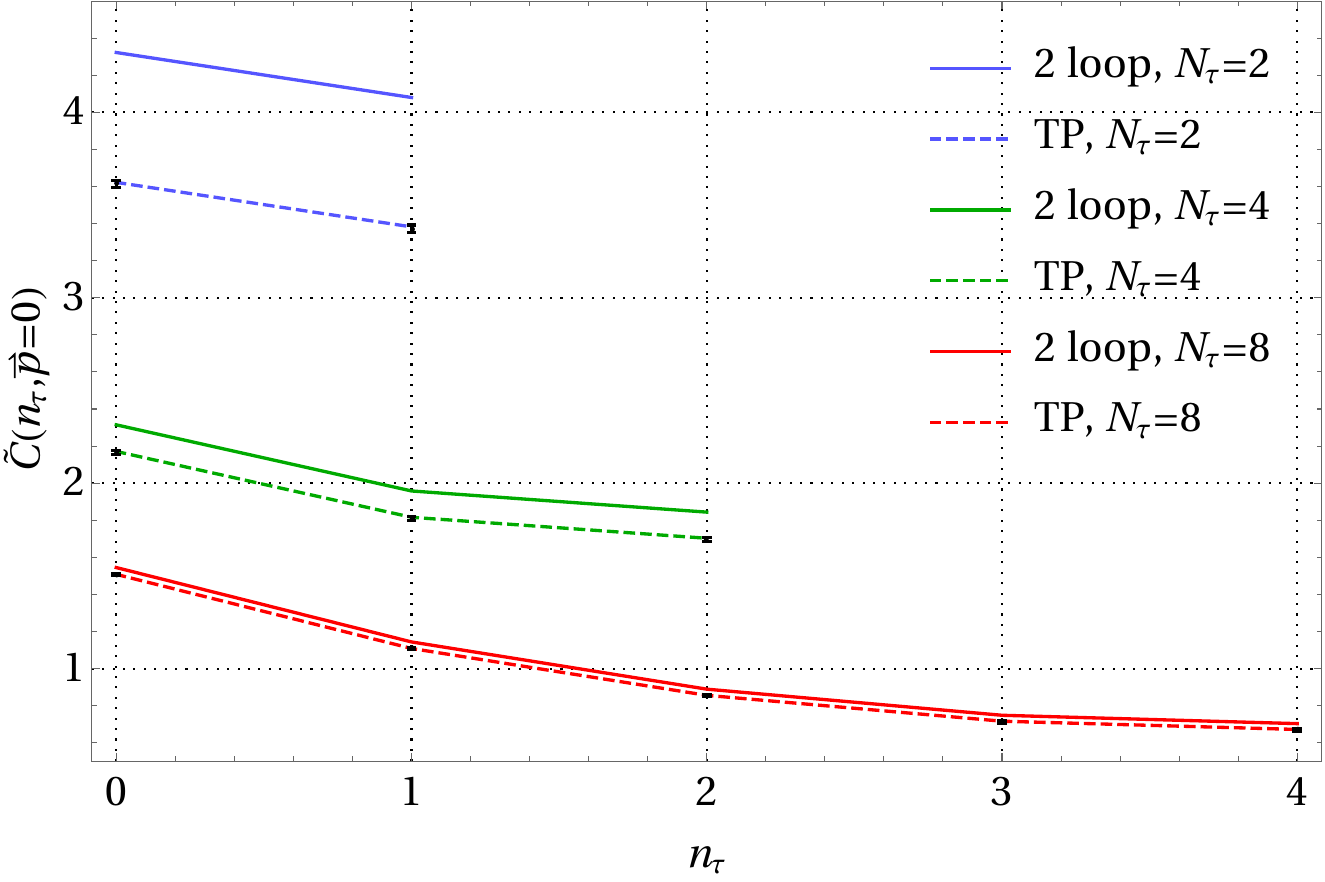}
\caption{Left: Thermoparticle spectral contribution for one-component $\phi^4$-theory in the symmetric phase, smaller $N_\tau$ implies
higher temperature. 
Right: Temporal correlator predicted from this spectral function, compared to data and perturbation theory.  From~\cite{Lowdon:2024atn} }
\label{fig2}
\end{figure}

The proposed method has recently been tested in $\phi^4$- theory for a one-component real scalar field~\cite{Lowdon:2024atn}.
Starting point is the spatial correlator averaged over the orthogonal directions, which is well described over all available
distance by a one-exponential screening mass fit (adapted to periodic boundary conditions)
\begin{align}
C(z) = A \, e^{-m_{\text{scr}}z} + A \, e^{-m_{\text{scr}}(aN_{s}-z)}.
\label{Cz_fit}
\end{align}
This implies a damping factor 
 \begin{align}
D_{m,\beta}(\bx) =  \alpha \, e^{-\gamma|\vec{x}|}, \quad\quad \alpha = 2 A \, am_{\text{scr}}, \  \gamma= m_{\text{scr}}-m.
\label{thermo_damping_scalar}
\end{align} 
The thermoparticle state contributes to the spectral function $\rho(\omega,\vec{p})$ as
\begin{align}
\rho_{\text{TP}}(\omega,\bp=0) = \epsilon(\omega)  \theta(\omega^{2}-m^{2}) \,  
\frac{4 \, \alpha \gamma  \sqrt{\omega^{2}-m^{2}}}{(\omega^{2}-m^2)^{2} + 2(\omega^{2}-m^{2})\gamma^{2}+\gamma^{4} },
\label{thermo_spec} 
\end{align}
which is shown in \fig\ref{fig2} (left) for different temperatures, $T=(aN_\tau)^{-1}$ and $0<T\lsi m$.. 
We observe the one-particle vacuum delta function gets 
damped and broadened by
collisions in the medium. As a non-perturbative test one can Fourier transform this spectral function to predict the temporal correlator,
and compare it to the corresponding lattice data, \fig\ref{fig2} (right).  A perfectly quantitative  description is obtained, from which we can conclude that the
continuous contributions (which are included in the full lattice correlator) are smaller than the statistical error bars in this temperature range. 
Note also that the non-perturbative prediction based on thermoparticles alone is far better than the perturbative prediction.

\section{Pseudo-scalar mesons in QCD}

Pseudo-scalar mesons are stable particles in QCD  and the considerations 
from the last section apply in a similar way. 
A complication is the richer vacuum spectrum of QCD. In particular, in the pion channel there are additional excitations like 
the $\pi^*$ and higher states, which on the lattice appear as stable particles unless multi-particle operators are used. 
In~\cite{Lowdon:2022xcl}, we analyse lattice data for spatial~\cite{Rohrhofer:2019qwq} and temporal~\cite{Rohrhofer:2019qal} meson
correlators obtained in simulations with $N_f=2$ domain wall fermions, i.e.~with good chiral symmetry even at finite lattice spacing.
The analysis is analogous to the one described for the scalar theory, except that in place of (\ref{Cz_fit}) we now use a 
two-state fit ansatz with terms of the same form for the $\pi$ and $\pi^*$, respectively. Such an ansatz provides a perfect description of the
spatial lattice correlator over its entire available range. 
\begin{figure}[t]
\centering
\includegraphics[width=0.45\textwidth]{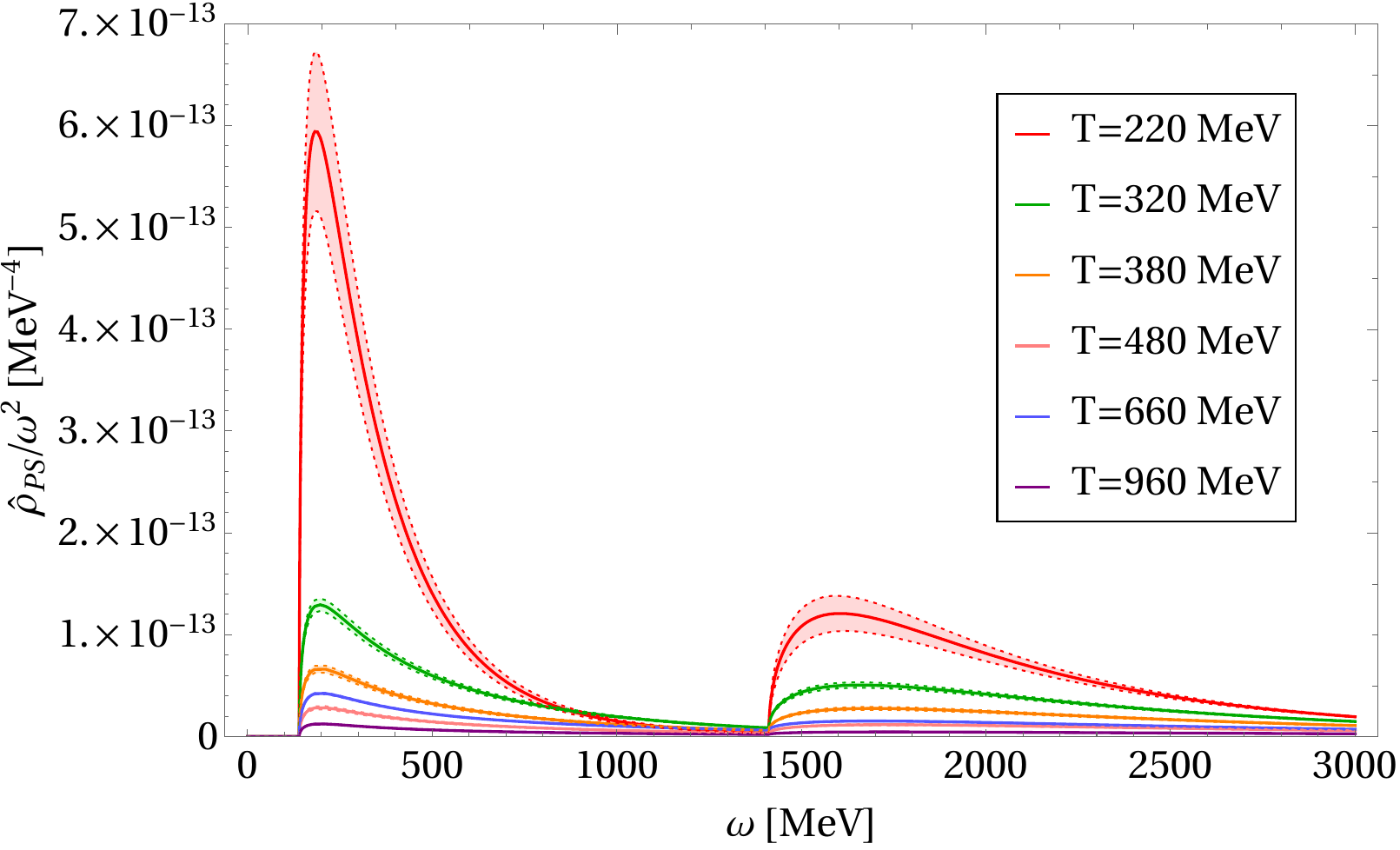}
\includegraphics[width=0.45\textwidth]{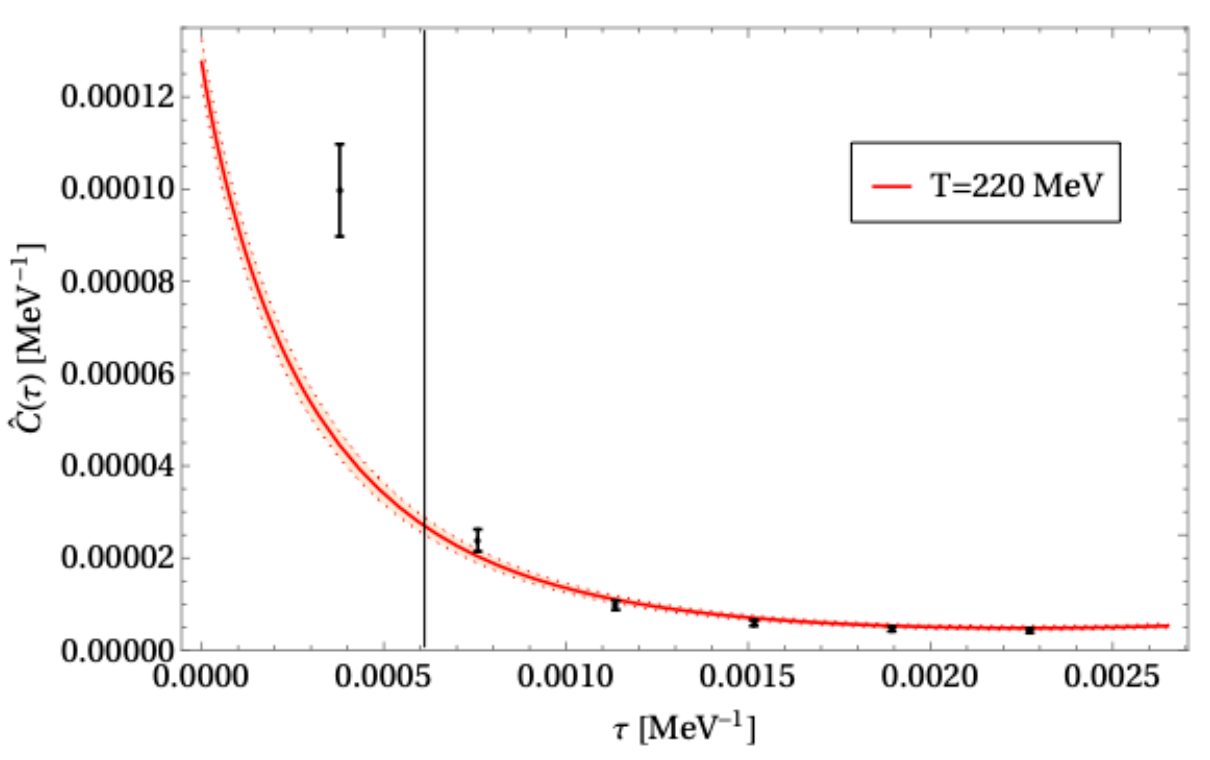}
\caption{Left: Spectral function for $\pi$ and $\pi^*$ thermoparticles. Right: Temporal correlator predicted from the spectral function compared
to lattice data~\cite{Rohrhofer:2019qal}. From~\cite{Lowdon:2022xcl}. }\label{fig3}
\end{figure}

The resulting spectral function for a range 
of temperature is shown in \fig\ref{fig3} (left). Again we observe clear thermoparticle peaks that can be identified with the $\pi$ and $\pi^*$, 
which broaden with increasing temperature until they lose their resonance-like structure. For the lowest temperature of $T=220$~MeV there
are lattice data also for the temporal correlator, with which the thermoparticle prediction can be compared, \fig\ref{fig3} (right).  Again the spectral
function provides a quantitatively accurate description of this correlator down to a distance scale $\sim m_{\pi^*}^{-1}$, which is marked by
the vertical line in the plot. Since the spectral function is constructed using information up to the $\pi^*$ scale only, it cannot properly represent
distances shorter than its inverse mass, for which knowledge of further excited states would be necessary. In principle, this is of course entirely possible
by calculating the same spatial correlator on finer lattices, but the thermal effects of interest are in infrared sector of the theory.

\begin{figure}[t]
\centering
\includegraphics[width=0.45\textwidth]{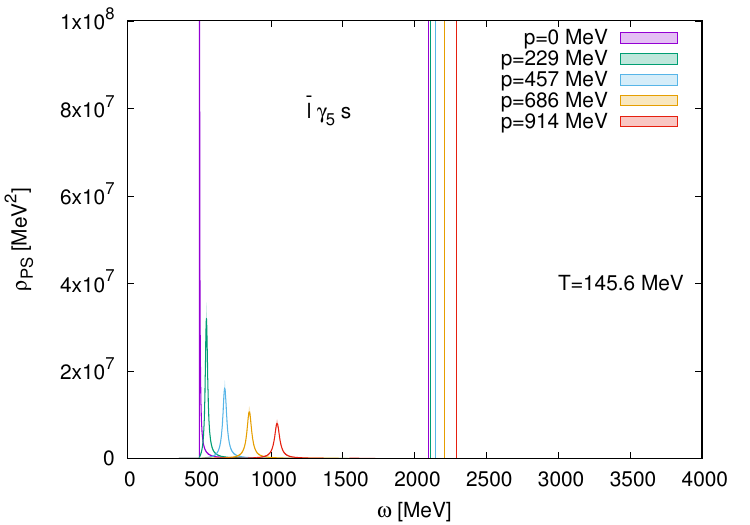}
\includegraphics[width=0.45\textwidth]{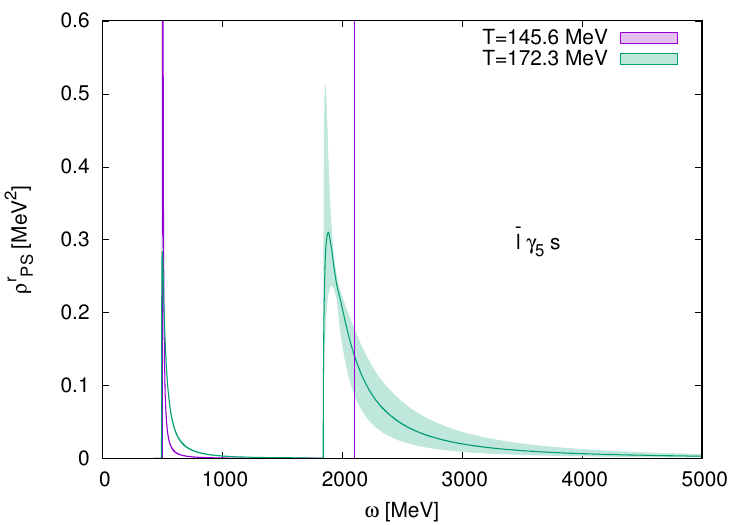}
\caption{Left: Thermoparticles in the kaon channel at various momenta. Right: Comparison of spectral functions slightly below
and above the chiral crossover. From \cite{Bala:2023iqu}.}\label{fig4}
\end{figure}

The analysis is further generalised in \cite{Bala:2023iqu} in two ways: firstly we now also include correlators for non-zero spatial momentum
to obtain the $\mathbf{p}$-dependence of the spectral functions. Secondly, we apply our analysis to  the strange-light and strange-strange 
pseudo-scalar channels in  $N_f=2+1$ QCD at $T=145.6$~MeV and $T=172.3$~MeV, i.e.~slightly below and above the chiral crossover temperature.
\fig\ref{fig4} (left) shows the spectral function obtained from a two-state fit with the kaon and its first excitation of the spatial correlators, now for different
momenta, with more damping and broadening the higher the momentum, as one might expect. We also note that the corresponding 
temporal correlator prediction deteriorates with increasing momentum, see \cite{Bala:2023iqu} for details. This is also expected,
because with increasing energy scale the continuous part of the spectrum (scattering states and Landau damping) starts to play an
inreasing role. 

\fig\ref{fig4} (right) compares the spectral function based on these two thermoparticle contributions at the lower and higher temperature.
In agreement with the pion situation, we see more damping effects as the temperature is increased, but no significant qualitative difference 
in the structure of the spectral function, which is then expected to set in at larger temperatures. 

\section{Conclusions}

Spectral functions in relativistic QFTs at finite temperature are strongly constrained by a combination of 
micro-causality, the KMS condition and the requirement to smoothly approach the K\"allen-Lehmann representation as $T\rightarrow 0$. 
We investigated a proposed decomposition of the thermal spectral density in terms of discrete thermoparticle contributions, corresponding
to thermally modified vacuum particles, 
and continuous conributions due to scattering, Landau damping and collective excitations. Neglecting the continuous parts allows to solve the inversion 
problem and gives quantitatively accurate descriptions for both the spatial \textit{and} temporal lattice correlators for temperatures of the order
of the lowest vacuum particle mass, which is relevant for the QCD chiral crossover. The persistence of resonance-like structures above 
the chiral transition is consistent with predictions based on an emergent chiral-spin symmetry~\cite{Rohrhofer:2019qal,Rohrhofer:2019qwq,Glozman:2022lda}.\\

\noindent
{\bf Acknowledgements:}
 The authors acknowledge support by the Deutsche Forschungsgemeinschaft (DFG, German Research Foundation) through the Collaborative Research Center CRC-TR 211 ``Strong-interaction matter under extreme conditions'' -- Project No. 315477589-TRR 211. O.~P.~also acknowledges support by the State of Hesse within the Research Cluster ELEMENTS (Project ID 500/10.006). 

\bibliographystyle{elsarticle-num} 
\bibliography{PS_refs}

\end{document}